\documentclass[11pt,a4paper]{article}
\usepackage{jcappub}

\usepackage[top=2.8cm, bottom=2.8cm, left=2.4cm, right=2.4cm]{geometry}
\usepackage[utf8]{inputenc} 
\usepackage[T1]{fontenc} 
\usepackage{amsmath,amssymb,lmodern} 
\usepackage{graphicx}
\usepackage{cancel}
\usepackage{color}
\usepackage{hyperref}

\title{Bosonic condensates in realistic supersymmetric GUT cosmic strings}

\author[a]{Erwan Allys}
\affiliation[a]{$\mathcal{G}\mathbb{R}\epsilon\mathbb{C}\mathcal{O}$, Institut d’Astrophysique de Paris, UMR 7095, 98 bis bd
 Arago, 75014 Paris, France \\ UPMC Universit\'e Paris 6 et CNRS,
 Sorbonne Universit\'es}
 \emailAdd{allys@iap.fr}
\arxivnumber{1505.07888}
\date{\today}

\abstract{
We study the realistic structure of $F$-term Nambu-Goto cosmic strings forming in a general supersymmetric Grand Unified Theory implementation, assuming standard hybrid inflation. Examining the symmetry breaking of the unification gauge group down to the Standard Model, we discuss the minimal field content necessary to describe abelian cosmic strings appearing at the end of inflation. We find that several fields will condense in most theories, questioning the plausible occurrence of associated currents (bosonic and fermionic). We perturbatively evaluate the modification of their energy per unit length due to the condensates. We provide a criterion for comparing the usual abelian Higgs approximation used in cosmology to realistic situations.}

\begin{document}
\maketitle

%%%%%%%%%%%%%%%%%%%%%%%%%%%%%%%%%%%%%%%%%%%%%%%%%%%%%%%%%%%%%%%%%

\section{Introduction}

Decade long improvements in experimental data led to the conclusion that only those Grand Unified Theories (GUT) involving some amount of Supersymmetry (SUSY) were acceptable \cite{Aulakh:2002zr,Fukuyama:2004ps,Fukuyama:2004xs,Aulakh:2004hm,Bajc:2004xe,Aulakh:2005mw,Bajc:2005qe,Aulakh:2003kg,Raby:2011jt}. The vacuum structure of these theories implies that they should have
produced topological defects during their successive steps of Spontaneous Symmetry Breaking (SSB),
including monopoles and cosmic strings \cite{Kibble:1980mv,Hindmarsh:1994re}, a phase of cosmic inflation being then necessary to dilute the former.
If we consider furthermore a $F-$term hybrid inflation scenario~\cite{Copeland:1994vg,Linde:1993cn,Dvali:1994ms,Lyth:1998xn,Kyae:2005vg,Mazumdar:2010sa,Davis:1999tk},
most of the SSB schemes lead to the formation of cosmic strings at the end of inflation \cite{Jeannerot:2003qv}.
So, constraining the string energy per unit length, e.g. through Cosmic Microwave Background (CMB) observations
\cite{Bouchet:2000hd,Bevis:2007gh,Ringeval:2010ca,Ade:2013xla,Brandenberger:2013tr} provides a general way to constrain GUT themselves. 

The structure of cosmic strings forming at the end of inflation has already been studied in details, see Refs.~\cite{Kibble:1982ae,Aryal:1987sn,Ma:1992ky,Davis:1996sp,Davis:1997bs,Ferreira:2002mg,Morris:1997ua,Davis:1997ny}. These works considered models where only the minimal field content necessary to form a string was introduced, and where the scale of formation of strings is the only dimensionful parameter. The aim of the present paper is to investigate the structure of cosmic strings in a realistic GUT context, i.e. starting from the complete GUT field content, and analyzing how the Higgs fields implementing the SSB scheme form the strings. This is done in $F$-terms models, focusing only on the bosonic part of the supermultiplets.

Considering a general SUSY GUT, we identify the minimal field content sufficient to describe the realistic string structure. This minimal structure involves all the fields which take non vanishing Vacuum Expectations Values (VEVs) at the end of the SSB scheme, and thus are singlet of the Standard Model (SM). The energy par unit length will be modified by this condensation of several Higgs fields in the core of the string. The additional fields also give natural candidates to carry bosonic currents \cite{Witten:1984eb,Peter:1992dw,Peter:1992nz,Peter:1993tm,Morris:1995wd}, and we expect their superpartner to carry zero-modes fermionic currents \cite{Witten:1984eb,Jackiw:1981ee,Weinberg:1981eu,Davis:1995kk,Ringeval:2000kz,Peter:2000sw}. Furthermore, the intercommutation process \cite{Shellard:1988ki,Laguna:1990it,Matzner:1988ky,Moriarty:1988qs,Moriarty:1988em,Shellard:1987bv} and the cusps evaporations \cite{Srednicki:1986xg,Brandenberger:1986vj,Gill:1994ic,Olum:1998ag,Bhattacharjee:1989vu} can be qualitatively modified due to this extra structure. The modification of any of these properties can have a major impact on the cosmological consequences of cosmic strings. In this paper, we give a complete description of the realistic microscopic structure of cosmic strings, and we perform a perturbative study of the modification of their energy per unit length from standard toy models. This gives a first step for the study of the other phenomena mentioned above.

For this purpose, we give an ansatz and boundary conditions for such a minimal structure, in the case of Nambu-Goto abelian strings. The conventions and normalizations chosen are such that if all the fields were to decouple from the string-forming Higgs, one would recover the standard abelian Higgs model. Two different classes of strings are discussed, only depending on how the Higgs fields of the GUT implement the SSB scheme, referred to as single and many-field strings. We perturbatively evaluate the modification of the energy per unit length from standard toy models due to this complex structure, taking into account the numerical factors appearing when one writes the complete formulation of the GUT.

In Sec.~\ref{GeneralModel}, we specify the SUSY GUT studied while setting the notation, and briefly describe its SSB scheme in parallel with the inflationary process. In Sec.~\ref{PartAbelianStrings}, using properties of the GUT reviewed in the first section, we present the abelian cosmic strings considered, discuss their minimal structure, and distinguish the two categories of strings. Finally, in Secs.~\ref{SingleFieldStrings} and~\ref{ManyFieldsStrings}, we propose ansätze and perform a perturbative study of their properties.

%%%%%%%%%%%%%%%%%%%%%%%%%%%%%%%%%%%%%%%%%%%%%%%%%%%%%%%%%%%%%%%%%

\section{Theoretical framework}
\label{GeneralModel}
%%%%%%%%%%%%%%%%%%%%%%%%%%%%%%%%%%%%%%%%%%%%%%%%%%%%%%%%%%%%%%%%%%

\subsection{Field content and superpotential}
\label{Superpotential}

We consider a general SUSY GUT associated with a gauge group $G$. Usual
examples are based on either SO(10) or SU(6). The spontaneous symmetry
breaking down to the SM will be implemented in
the context of a $F$-term hybrid inflation
\cite{Copeland:1994vg,Linde:1993cn,Dvali:1994ms,Lyth:1998xn,Kyae:2005vg,Mazumdar:2010sa,Davis:1999tk}. The field content of the theory includes a set of chiral supermultiplets
and a gauge supermultiplet associated with the generators of $G$.
We restrict ourselves to the bosonic sector of the
model, and so only write down the scalar part of the chiral
supermultiplets. As we study a $F$-term theory, we assume the
$D$-terms are identically zero,
with no Fayet-Iliopoulos terms~\cite{Martin:1997ns}. It will add some constraints to the fields.

In order to implement hybrid inflation, it is a necessity to have at
least two fields $ \mathbf{\Sigma}$ and $\overline{ \mathbf{\Sigma}}$ in complex conjugate
representations, which have a coupling term with the inflaton in the
superpotential \cite{Copeland:1994vg,Linde:1993cn,Dvali:1994ms}. The inflaton is assumed to be a chiral supermultiplet
of scalar component $S$, singlet under $G$. In order to reach the SM
symmetry and have a phase of inflation washing away monopoles, one
needs to have more than one SSB step; this means other fields
should be present. The other chiral supermultiplets are
denoted $\mathbf{\Phi}_I$ for the fields in real representations, and $\mathbf{\Phi}_i$
and $\mathbf{\Phi}_{\bar{\imath}}$ for the fields in complex representations.

We assume the most general superpotential taking into
account all the above chiral supermultiplets, supplemented by
a specific term for the inflaton $S$ to implement hybrid inflation. It is,
assuming explicit summation on all repeated indices, \cite{Martin:1997ns}
\begin{align}
\label{EqSuperpotential}
W=&~ m_\Sigma \mathbf{\Sigma} \overline{ \mathbf{\Sigma}} + \frac{1}{2} m_{IJ} \mathbf{\Phi}_I \mathbf{\Phi}_J + m_{i\bar{\jmath}} \mathbf{\Phi}_i \mathbf{\Phi}_{\bar{\jmath}} 
+ \eta_x \mathbf{\Sigma} \overline{ \mathbf{\Sigma}} \mathbf{\Phi}_x \nonumber \\
& + \beta_{xy} \mathbf{\Sigma} \mathbf{\Phi}_x \mathbf{\Phi}_y + \bar{\beta}_{xy} \overline{ \mathbf{\Sigma}} \mathbf{\Phi}_x \mathbf{\Phi}_y 
+ \frac{1}{3}\lambda_{xyz} \mathbf{\Phi}_x \mathbf{\Phi}_y \mathbf{\Phi}_z + \kappa S ( \mathbf{\Sigma}\overline{ \mathbf{\Sigma}} - M^2),
\end{align}
the last term actually implementing hybrid inflation. In this
equation, the label $x$, $y$ and $z$ can be either $I$, $i$ or
$\bar{\imath}$. All the coefficients which appear in addition to the fields
are complex constants, and $\beta_{xy}$, $\bar{\beta}_{xy}$ and
$\lambda_{xyz}$ are totally symmetric in their indices. The constants
$\kappa$, $M$, $m_\Sigma$ and the diagonal elements of the mass
matrices can be set real and positive by redefinition of the phases of
the fields. Depending on the explicit choices of representations, the
coefficients $m_{IJ}$, $m_{i\bar{\jmath}}$, $\eta_x$, $\beta_{xy} $,
$\bar{\beta}_{xy} $ and $\lambda_{xyz}$ will be non-zero only when they allow to build gauge singlets. We did not include terms like
$\mathbf{\Sigma}\mathbf{\Sigma}\mathbf{\Phi}$ and
$\mathbf{\overline{\Sigma}}\mathbf{\overline{\Sigma}}\mathbf{\Phi}$ for the sake of simplicity; their contribution to the macroscopic structure of the string is shortly discussed in Sec. \ref{ManyFieldsModifEnergy}, and found similar to that of terms like $\mathbf{\Sigma}\mathbf{\overline{\Sigma}}\mathbf{\Phi}$.

In this work, we chose to consider separately the GUT and inflation parts, being as realistic as possible for the former and setting up the simplest possible model for the latter. This is the approach found also e.g. in \cite{Cacciapaglia:2013tga}. This simple inflation term is sufficient to reproduce the standard inflation phenomenology. Besides, different kinds of terms such as $S^2$, $S^3$ or $S \mathbf{\Phi}^2$ generate mass or quartic terms for the inflaton in the scalar potential and can thus spoil the inflation. They must therefore be considered with care \cite{Cacciapaglia:2013tga}.

In the example of a SO(10) GUT, the fields $ \mathbf{\Sigma}$ and $\overline{\mathbf{\Sigma}}$
are often taken to transform as the $\mathbf{126}$ and
$\mathbf{\overline{126}}$ representations, which are the lowest dimensional complex conjugate
representations which are safe, i.e. permitting $R$-parity conservation at low energy to ensure proton stability
\cite{Martin:1992mq}. The other fields used to implement the SSB can
be for example a $\mathbf{210}$ and a $\mathbf{10}$
\cite{Aulakh:2003kg}, or two $\mathbf{54}$, two $\mathbf{45}$, and two
$\mathbf{10}$ representations \cite{Jeannerot:1995yn}. The whole expression of a SO(10) model, with the normalizations and conventions of the present paper, is given in Ref.~\cite{Allys:2015kge}.

%%%%%%%%%%%%%%%%%%%%%%%%%%%%%%%%%%%%%%%%%%%%%%%%%%%%%%%%%%%%%%%%%%

\subsection{Lagrangian of the bosonic sector}

Let us write down the general form for the Lagrangian of
the bosonic sector, while setting the notation. In
the following, we take the signature of the metric to be $+2$, and
label with latin indices $a$, $b$, $\dots$ the generators of G, with gauge coupling constant
$g$. The kinetic part of the Lagrangian is thus (with implicit summations on $i$, $\bar{i}$ and $I$)
\begin{multline}
\label{Kinetic}
K =- (D_\mu \mathbf{\Phi}_{\bar{\imath}})^\dagger (D^\mu \mathbf{\Phi}_{\bar{\imath}})
 - ( D_ \mu\mathbf{\Phi}_i)^\dagger( D^\mu\mathbf{\Phi}_i) 
 - (D_\mu\mathbf{\Phi}_I)^\dagger(D^\mu\mathbf{\Phi}_I) \\
 -(D_\mu \overline{ \mathbf{\Sigma}})^\dagger (D^\mu \overline{ \mathbf{\Sigma}}) 
 - ( D_\mu \mathbf{\Sigma})^\dagger  ( D^\mu \mathbf{\Sigma}) 
  - (\nabla_\mu S)^* (\nabla^\mu S) -\frac{1}{4}F_{\mu \nu}^a F^{a \mu \nu},
\end{multline}
with
\begin{equation}
\label{CovDev}
D_\mu X=(\nabla_\mu - i g A_\mu^a \tau^a_X)X,
\end{equation}
$\tau^a_X$ being the relevant operators encoding the action of the
generator labeled by $a$ on the field $X$. From now on, we will
denote by $X$ a generic scalar field when unspecified, 
i.e. $X\in \{ \mathbf{\Sigma}, \overline{ \mathbf{\Sigma}}, \mathbf{\Phi}_I, \mathbf{\Phi}_i, \mathbf{\Phi}_{\bar{\imath}},S\}$.
%The gauge field can also be written $A^\mu=A_\mu^a \tau^a$, but it is
%important to take care that the field over which this operator acts
%is not written explicitly in this expression, although the operator
%actually depends of this field.
The strength tensor is defined in the usual way,
\begin{equation}
F_{\mu\nu}^a=\nabla_\mu A_{\nu}^a-\nabla_\nu A_{\mu}^a+g
f^a{}_{bc}A_\mu^b A_\nu^c,
\end{equation}
with $f^a{}_{bc}$ the structure constants of $G$.

The potential term is constructed from the $F$-term, obtained by
taking the derivative of the superpotential with respect to the chiral
supermultiplets
\begin{equation}
F_X=\frac{\partial W}{\partial X}.
\end{equation}
This yields the terms
\begin{equation}
\label{F-terms}
\begin{array}{l}
\mathbf{F}_{\Sigma}=\overline{ \mathbf{\Sigma}} (m_\Sigma +\eta_x \mathbf{\Phi}_x + \kappa S)
+\beta_{xy} \mathbf{\Phi}_x \mathbf{\Phi}_y,\\ 
\mathbf{F}_{\bar{\Sigma}}= \mathbf{\Sigma}(m_\Sigma +\eta_x
\mathbf{\Phi}_x + \kappa S) +\bar{\beta}_{xy} \mathbf{\Phi}_x \mathbf{\Phi}_y,\\
 \mathbf{F}_{\Phi_I}=m_{IJ}\mathbf{\Phi}_J + \eta_I
\overline{ \mathbf{\Sigma}} \mathbf{\Sigma} + 2 \beta_{Iy} \mathbf{\Sigma} \mathbf{\Phi}_y +
2 \bar{\beta}_{Iy}\overline{ \mathbf{\Sigma}}\mathbf{\Phi}_y + \lambda_{Iyz}\mathbf{\Phi}_y
\mathbf{\Phi}_z,\\
 \mathbf{F}_{\Phi_i}=m_{i\bar{\jmath}}\mathbf{\Phi}_{\bar{\jmath}} + \eta_i
\overline{ \mathbf{\Sigma}} \mathbf{\Sigma} + 2 \beta_{iy} \mathbf{\Sigma} \mathbf{\Phi}_y +
2 \bar{\beta}_{iy}\overline{ \mathbf{\Sigma}}\mathbf{\Phi}_y + \lambda_{iyz}\mathbf{\Phi}_y
\mathbf{\Phi}_z,\\ 
\mathbf{F}_{\Phi_{\bar{\imath}}}=m_{j\bar{\imath}}\mathbf{\Phi}_j + \eta_{\bar{\imath}}
\overline{ \mathbf{\Sigma}} \mathbf{\Sigma} + 2 \beta_{\bar{\imath}y} \mathbf{\Sigma} \mathbf{\Phi}_y + 2 \bar{\beta}_{\bar{\imath}y}\overline{ \mathbf{\Sigma}}\mathbf{\Phi}_y +
\lambda_{\bar{\imath}yz}\mathbf{\Phi}_y \mathbf{\Phi}_z,
\\ F_{S}=\kappa
(\overline{ \mathbf{\Sigma}} \mathbf{\Sigma}-M^2),
\end{array}
\end{equation}
$F_X$ being in the conjugate representation of $X$. These $F$-terms are discussed in more details in Sec.~\ref{LagEOM}. The scalar
potential is finally obtained from these terms through
\begin{equation}
\label{potential}
V=\sum_X \mathbf{F}_X^\dagger \mathbf{F}_X\equiv\sum_X V_X,
\end{equation}
where $V_X>0$ and $V>0$, and where we use additional bold symbols for the $F$-terms to remind that they are not singlet of the gauge group in general. Note that we did not include in $V$ the $D$-term contribution, since these terms identically vanish in a $F$-term scenario, and thus play no role in the dynamical study of the fields. However, they are indeed taken into account by imposing some constraints which are discussed in Sec.~\ref{PartAnsatzBCSingle} and~\ref{IntroManyFields}.

Finally, the full Lagrangian density is derived from these two
terms
\begin{equation}
\label{L=K-V}
\mathcal{L}=K-V,
\end{equation}
i.e. by adding Eqs. (\ref{Kinetic}) and (\ref{potential}).

%%%%%%%%%%%%%%%%%%%%%%%%%%%%%%%%%%%%%%%%%%%%%%%%%%%%%%%%%%%%%%%%%%

\subsection{Hybrid inflation and SSB scheme}
\label{HI&SSB}

It has been shown in \cite{Jeannerot:2003qv} that the formation of cosmic strings at the end of an hybrid inflation phase is essentially unavoidable during the final stage of a SUSY GUT symmetry breaking. 
It is this category of strings that we discuss in the present paper. In such models, all the monopoles which may have been produced in a previous phase are washed out during inflation.

The matter content we introduced previously implements
the SSB scheme down to the SM in at least two steps :
\begin{equation}
G
\overset{\langle\Phi_x\rangle}{\relbar\joinrel\relbar\joinrel\cdots\joinrel\longrightarrow}
G' \overset{\langle\Sigma\rangle\langle\Phi_{x'}\rangle}
{\relbar\joinrel\relbar\joinrel\relbar\joinrel\longrightarrow}
G_{\text{SM}} \times \mathbf{Z}_2.
\end{equation}
In many cases, the symmetry
that is broken at the last step contains U(1)$_{B-L}$, but since there are no
general constraints about it, it is better to leave it arbitrary.

All the non vanishing VEVs after the end of inflation have to be
singlet under the SM gauge group, otherwise the vacuum would
have non vanishing quantum numbers under this symmetry.
We also assume there is no symmetry restoration, i.e. all fields acquiring a non-zero VEV at a given stage keep it non vanishing at later stages. So, we can restrict the study of the SSB scheme to SM singlets only.

At the onset of inflation, we can assume an initially very large
value for the inflaton $S$ in comparison with all the other
fields, as is expected for chaotic inflation. To minimize $V_\Sigma
\sim |\kappa \overline{ \mathbf{\Sigma}} S|^2$ and $V_{\bar{\Sigma}} \sim |\kappa
 \mathbf{\Sigma} S|^2$, the fields $ \mathbf{\Sigma}$ and $\overline{ \mathbf{\Sigma}}$ must take a
vanishing VEV. 
The terms $\beta_{xy} \mathbf{\Phi}_x \mathbf{\Phi}_y$ must be in the same
representation as $\overline{ \mathbf{\Sigma}}$ for the terms $\beta_{xy} \mathbf{\Phi}_x \mathbf{\Phi}_y \mathbf{\Sigma}$ to be scalars. So, if they take a VEV before the end of inflation, all the symmetries
broken by $\overline{ \mathbf{\Sigma}}$ after inflation would already be broken at this current step, which is contrary to our assumptions. Thus, they cannot take a non-zero VEV before the end of inflation. 
This property and
the same reasoning for $\bar{\beta}_{xy} \mathbf{\Phi}_x \mathbf{\Phi}_y$ show that
$V_\Sigma=0$ and $V_{\bar{\Sigma}}=0$ before the end of inflation.
This also ensures that all the terms implying the inflaton $S$ 
identically vanish in the potential at tree level.

Finally, we
can assume that at the onset of inflation, the VEVs verify that all
the potential terms except $V_S$ are zero; it is indeed the
global minimum for the potential taking into account the constraint 
on $V_S$. Different field configurations give this minimal value, 
including that with all the fields having a vanishing value\footnote{We assume that other configurations can exist, since the potential conditions 
give fourth order polynomial equations in the fields.}. 
So, it gives several sets of solutions
\begin{equation}
\label{VEV0}
\{\langle \mathbf{\Sigma}_{\scriptscriptstyle{(-)}}\rangle=0,\langle\overline{ \mathbf{\Sigma}}_{\scriptscriptstyle{(-)}}\rangle=0,\langle \mathbf{\Phi}_{x,{\scriptscriptstyle{(-)}}}\rangle\}_{\delta},
\end{equation}
the index $(-)$ meaning that we consider the set of VEVs before the
end of inflation, while $\delta$ labels the different sets
themselves.

At the end of inflation, the fields reach a global
minimum for the potential, meaning all $F$-terms contributions must independently
vanish. Setting $V_S=0$ then implies
\begin{equation}
\label{SigmaInf}
\langle \mathbf{\Sigma}\rangle\langle\overline{ \mathbf{\Sigma}}\rangle=M^2,
\end{equation}
and the fields $ \mathbf{\Sigma}$ and $\overline{ \mathbf{\Sigma}}$ take non zero VEVs as expected.
The vanishing conditions of all the other potential terms finally
give several sets of solutions
\begin{equation}
\label{VEV1}
\{\langle \mathbf{\Sigma}_{\scriptscriptstyle{(+)}}\rangle,\langle\overline{ \mathbf{\Sigma}}_{\scriptscriptstyle{(+)}}\rangle,S_{\scriptscriptstyle{(+)}},\langle\mathbf{\Phi}_{x,{\scriptscriptstyle{(+)}}}\rangle\}_{\delta^\prime},
\end{equation} 
the index $(+)$ denoting that we consider the set of VEVs after the end
of inflation, while $\delta^\prime$ labels the different sets of solution.

During inflation, the potential for the inflaton yields $V=V_0 + \text{quant.corr.}$, where $V_0=\kappa^2 M^4$ is the value of the potential at tree-level. Due to the quantum corrections, the value of the inflaton field will then slowly roll, until it meets its critical value, thus ending inflation (see Ref.\cite{Cacciapaglia:2013tga} for an explicit example). 

As we study strings in models which are relevant in a particle physics and cosmological point of view, we assume that the superpotential and field content used imply at least one appropriate SSB scheme. For these schemes, the set of VEVs at the end of inflation defines the SM gauge group, no harmful topological defects are produced, and the stability of the inflationary valley is ensured (see the associated discussion\footnote{It was also
argued in Ref.~\cite{Cacciapaglia:2013tga} that one should
verify that the set of VEVs before the end of inflation is close in
field space to only one set of VEVs after the end of inflation, since otherwise the possibility that the fields take different sets of VEVs at the end of
inflation could create domain
walls.} in Ref.~\cite{Cacciapaglia:2013tga}). We consider from now on such a SSB scheme and the associated non vanishing VEVs, those being described as in Eqs.~(\ref{VEV0})
and (\ref{VEV1}) without considering the labels $\delta$ and $\delta '$
anymore, i.e. explicitly assuming a specific set in each ensemble.

Such a SSB scheme in parallel with the inflationary process in a given SO(10) GUT is described in Ref.~\cite{Allys:2015kge}.

%%%%%%%%%%%%%%%%%%%%%%%%%%%%%%%%%%%%%%%%%%%%%%%%%%%%%%%%%%%%%%%%%%

\subsection{Description with the restricted representations}
\label{RestrictedRep}

As we work with large dimensional representations, it is useful to consider
their branching rules to simplify the description of these fields. Indeed, we only need to work with fields behaving as singlets under the SM
gauge group to describe the whole SSB scheme. For instance, the
\textbf{210} representation of SO(10) contains three such restricted
representations, included in its 
(\textbf{1},\textbf{1},\textbf{1}), (\textbf{1},\textbf{1},\textbf{15}) 
and (\textbf{1},\textbf{3},\textbf{15}) representations of its decomposition 
under the Pati-Salam group [SU(2)$_L\times$SU(2)$_R\times$SU(4)] \cite{Slansky:1981yr}.
The \textbf{126} and $\mathbf{\overline{126}}$ representations only have one such sub-representation, contained respectively
in their (\textbf{1},\textbf{3},\textbf{10}) and (\textbf{1},\textbf{3},$\mathbf{\overline{10}}$)
restricted representations under Pati-Salam.

Let us consider such a VEV singlet. For
a given field $\mathbf{\Phi}_x$, we ascribe an index $\alpha$ to describe the
different sub-representations transforming trivially under the SM, and write the associated VEVs as
%\footnote{To write the singlets of the SM in this form, we need to assume that their are not charged under any non abelian continuous symmetry which commute with the SM gauge group. It is a very standard assumption when studying GUTs of rank $5$ or $6$. For higher rank groups, it is sufficient to focus on the late part of the SSB scheme.}
\begin{equation}
\label{defsubmultiplet}
\langle \mathbf{\Phi}_{x,\alpha} \rangle = \phi_{x,\alpha} \left(x^\mu\right) {\langle \mathbf{\Phi}_{x,\alpha}\rangle}_0,
\end{equation}
without implicit summation, and where $\phi_{x,\alpha}$ is a complex function of space-time and ${\langle
 \mathbf{\Phi}_{x,\alpha}\rangle}_0$ a constant normalized vector in representation
space (${\langle
 \mathbf{\Phi}_{x,\alpha}\rangle}_0^\dagger{\langle
 \mathbf{\Phi}_{x,\alpha}\rangle}_0=1$). Using these notations, the complete part of the field which is singlet under the SM
can be written as the combination
\begin{equation}
\label{SubMultiplet}
\langle \mathbf{\Phi}_x \rangle = \sum_\alpha \phi_{x,\alpha}(x^{\mu}) {\langle \mathbf{\Phi}_{x,\alpha}\rangle}_0.
\end{equation}
This procedure reduces the fields description to only a few complex
functions.

%%%%%%%%%%%%%%%%%%%%%%%%%%%%%%%%%%%%%%%%%%%%%%%%%%%%%%%%%%%%%%%%%%
%%%%%%%%%%%%%%%%%%%%%%%%%%%%%%%%%%%%%%%%%%%%%%%%%%%%%%%%%%%%%%%%%%
\section{Abelian cosmic strings}
\label{PartAbelianStrings}
%%%%%%%%%%%%%%%%%%%%%%%%%%%%%%%%%%%%%%%%%%%%%%%%%%%%%%%%%%%%%%%%%%
\subsection{Strings studied}

Let us focus on the strings created at the last step of
SSB of the GUT, lowering the rank of the gauge group by one unit and ending
the inflation phase. Several kinds of strings can
appear, depending on the quotient group $H\sim G^\prime / G_{\text{SM}}$. As the SSB lowers 
the rank of the group, $H$ must contain at least one U(1) subgroup, and we will focus on
the Nambu-Goto abelian strings which form at this step, associated with this abelian
generator. When $H$ is larger than U(1), non abelian strings
could also form \cite{Aryal:1987sn,Ma:1992ky,Davis:1996sp}. However, and since we want to constraint all GUTs, we restrict attention to the minimal U(1) case, any other kind of strings tightening the constraints. We will denote
$\text{U}(1)_{\text{str}}$ this particular subgroup and
$\tau^{\text{str}}$ the associated generator.

These strings cannot be connected to monopoles. As shown in Ref. \cite{Vilenkin:1982hm}, the strings are stable with respect to breaking into monopoles when the scale of formation of these monopoles is higher than the scale of formation of the strings. It is the case here, as we assumed that only strings form at the SSB considered. So, these strings could connect only to pre-existing monopoles, and those have, by construction, already been washed away during the inflation phase.

In what follows, we use a set of cylindrical coordinates $(r,\theta,z,t)$ based on the location 
of the string, and taken to be locally aligned along the $z$-axis at $r=0$. 
We also focus on strings with fields functions
of $r$ and $\theta$ only, and
consider uniquely the bosonic structure of the string. It means that we will not consider in this paper any currents in the core of the string. This possibility will however be discussed in the following section.

%%%%%%%%%%%%%%%%%%%%%%%%%%%%%%%%%%%%%%%%%%%%%%%%%%%%%%%%%%%%%%%%%%
\subsection{Minimal structure}
\label{PartMinimalStructure}

We have to determine which sub-representations are sufficient
to describe the structure of the string. On the one hand, the fields take at infinity the non vanishing VEVs defining the SM symmetry, so all the restricted representations non singlet under this symmetry take vanishing values far from the string. On the other hand, the sub representations non singlet under the SM appear at least in a quadratic form in the potential. Indeed, a potential term containing only one such field would be charged under the SM. Both these results imply that an ansatz where all the sub-representations charged under the SM take an identically vanishing value is solution of the equations of motion with the boundary conditions at infinity.

To understand in another way this ansatz, one can consider the static configuration which minimizes the potential at the center of the string. Such a configuration is given by the set of non vanishing VEVs singlet under the SM defining $G^\prime$ in the SSB scheme, see Sec.~\ref{HI&SSB}. These VEVs are not charged under U(1)$_{\text{str}}$, which would otherwise already be broken at this step. Using arguments discussed e.g. in Ref.~\cite{Witten:1984eb}, we expect fields in the string to take intermediate values between the configurations which minimize the potential at the center of the string and at infinity, depending on the competition between kinetic and potential terms. Then, as both these configurations only contain singlets of the SM, one expects an ansatz where all the other fields take an identically vanishing value. 

This particular ansatz, which we consider in the following, is what we define as the minimal structure. As the configurations which minimize the potential at the center of the string and at infinity are \emph{a priori} different since they are solutions of two different quartic polynomial equations, all the fields taking non vanishing values at the last step of SSB condense in the string, and thus have to be taken into account. Note that a complete study of the stability of such an ansatz should be considered in each given model.

The minimal structure for cosmic strings, where several bosonic fields condense in the string, should modify most of the properties of these objects. Some of these properties are described below. For instance, it becomes mandatory to describe the microscopic structure of the strings by taking into account the condensation of these additional Higgs fields. This then permits to evaluate the modification of the energy per unit length due to this complex structure. As most of the observational constraints on this macroscopic parameter are rather stringent, every sizable modification of it in a given model could rule out this model.

Several other properties of cosmic strings will be modified by the fact that the actual structure is more involved than that of the toy models usually considered. First, the condensation of several Higgs fields in the core of the string gives natural candidates to carry bosonic currents \cite{Witten:1984eb,Peter:1992dw,Peter:1992nz,Peter:1993tm,Morris:1995wd}. Also, and as we work in a SUSY framework, we can expect the superpartners of these Higgs fields to build fermionic currents via their zero modes \cite{Witten:1984eb,Jackiw:1981ee,Weinberg:1981eu,Davis:1995kk,Ringeval:2000kz,Peter:2000sw}. This more complex structure can also qualitatively modify the intercommutation process \cite{Shellard:1988ki,Laguna:1990it,Matzner:1988ky,Moriarty:1988qs,Moriarty:1988em,Shellard:1987bv}, which has a tremendous impact on the temporal evolution of the cosmological string network, and thus on the consequences on the CMB \cite{Kibble:1976sj,Kibble:1980mv,Copeland:1991kz,Austin:1993rg,Martins:1996jp,Martins:2000cs}. Furthermore, the link made between the fields forming the string and the particle physics model used allows a more detailed examination of the cusps evaporation phenomenon \cite{Srednicki:1986xg,Brandenberger:1986vj,Gill:1994ic,Olum:1998ag,Bhattacharjee:1989vu}. A modification of each of these properties can have major consequences on the cosmological implications of cosmic strings.

%%%%%%%%%%%%%%%%%%%%%%%%%%%%%%%%%%%%%%%%%%%%%%%%%%%%%%%%%%%%%%%%%%

\subsection{Equation of state, toy model limit}
\label{ToyModel}

Since we explicitly assume currentless strings, nothing in the configuration we are 
interested in can depend on the internal string worldsheet coordinates,
here locally $z$ and $t$. We have 
\begin{equation}
T^\mu_\nu=-2g^{\mu \alpha}\frac{\delta \mathcal{L}}{\delta
 g^{\alpha\nu}} + \delta^\mu_\nu \mathcal{L},
\end{equation}
yielding $T^{tt}=-T^{zz}=\mathcal{L}$. Then
\begin{equation}
\label{Nambu-Goto}
{\displaystyle U = 2\pi \int r\text{d}r~ T^{tt}=-2\pi \int r \text{d}r~ T^{zz}=T},
\end{equation}
i.e. the Nambu-Goto equation of state, Lorentz-invariant along the worldsheet. 
Thus, the only parameter of interest is the energy per unit length $U$ defined in Eq.~(\ref{Nambu-Goto}). This parameter, frequently denoted by $\mu$ in the cosmology literature, is directly constrained from, e.g., CMB observations: from Planck and WMAP data, there is a constraint of $G\mu/c^2 < 3.2\times 10^{-7}$ at $95\%$ confidence level for the abelian Higgs model \cite{Ade:2013xla}. Thus, every modification of $U$ due to the realistic structure of the strings will have to be compared to the already stringent observational constraints, and could rule out the associated model.

To translate the standard abelian Higgs model to a $F$-term SUSY formalism, three fields are necessary, $\Sigma$, $\overline{\Sigma}$ and $S$, the first two fields having opposite U(1) charges, the last one being uncharged. The superpotential is 
\begin{equation}
W=\kappa S \left( \overline{\Sigma} \Sigma -M^2\right).
\end{equation}
Assuming that $S$ identically vanishes, it yields as expected the standard U(1) string model for $\mathbf{\Sigma}$ and $\overline{\mathbf{\Sigma}}$ (see for example \cite{Davis:1997bs,Aryal:1987sn,Peter:1992dw}), with\footnote{Note that if one is more familiar with the model containing a single kinetic term of the form $(D_\mu \overline{\Sigma})(D^\mu \Sigma)$, usual for non supersymmetric theories, a link between both models can be easily performed. For this, it is sufficient to introduce new variables of the form $\tilde{\Sigma}=\sqrt{2}\Sigma$, $\tilde{M}=\sqrt{2}M$, and $\tilde{\kappa}=\kappa/2$. Indeed, one sees that the superpotential in term of the new variables remains the same, while the factor 2 of the kinetic term will disappear. This is only valid in a $F$-term scenario, where $\Sigma$ and $\overline{\Sigma}$ are complex conjugate due to the $D$-term condition (see \ref{PartAnsatzBCSingle}).}.
\begin{equation}
\mathcal{L}=-(D_\mu \overline{\Sigma})^\dagger(D^\mu \overline{\Sigma}) -(D_\mu \Sigma)^\dagger (D^\mu \Sigma)-\frac{1}{4} F_{\mu\nu}F^{\mu\nu}-\kappa^2 \left| \Sigma \overline{\Sigma}-M^2 \right| ^2.
\end{equation}

We recover explicitly this limit from the realistic model when all the parameters but $\kappa$ and $M$ go to zero, and considering only $\text{U}(1)_{\text{str}}$ defined in the previous part. Note that it is the case due to the previous normalizations and conventions, see Sec.~\ref{SingleFieldStrings} for the details. This well defined limit must be verified when using results from the abelian Higgs model, since important GUT numerical factors can appear. For this well known problem, we have $\Sigma \sim \overline{\Sigma} \sim M$, a characteristic radius for $\Sigma$ of $(\kappa M)^{-1}$, and a characteristic energy per unit length $U_0 \sim M^2$ \cite{Aryal:1987sn,Peter:1992dw,Hindmarsh:1994re,Allys:2015kge}.

%%%%%%%%%%%%%%%%%%%%%%%%%%%%%%%%%%%%%%%%%%%%%%%%%%%%%%%%%%%%%%%%%%
\subsection{Two classes of strings}
\label{PartTwoClasses}

We assume for the entire paper that $ \mathbf{\Sigma}$ and $\overline{\mathbf{\Sigma}}$ only have one sub-representation singlet of the SM, charged under U(1)$_{\text{str}}$. If it was not the case, we could treat the additional representations, charged or not under U(1)$_{\text{str}}$, in the same manner than the additional fields $\mathbf{\Phi}$. 

We distinguish two classes of strings. For the first kind, no other fields are charged under U(1)$_{\text{str}}$. We call these 
strings single-field strings, since only one field (and its conjugate) is directly coupled 
to the gauge field. They are discussed in Sec. \ref{SingleFieldStrings}. 
For the second case, other fields can be charged under 
$\text{U}(1)_{\text{str}}$, and we naturally call them many-field strings, see Sec. \ref{ManyFieldsStrings}. This property only depends on the field content of the GUT and how it permits to implement the SSB scheme.

In the case of a single-field string, no $\beta$-couplings, e.g. in $\mathbf{\Sigma \Phi \Phi}$, can appear in the superpotential between the restricted representations singlet under the SM, since such terms would be charged under $\text{U}(1)_{\text{str}}$. It is not anymore the case with a many-field string.

%%%%%%%%%%%%%%%%%%%%%%%%%%%%%%%%%%%%%%%%%%%%%%%%%%%%%%%%%%%%%%%%%%

\section{Single-field strings}
\label{SingleFieldStrings}
%%%%%%%%%%%%%%%%%%%%%%%%%%%%%%%%%%%%%%%%%%%%%%%%%%%%%%%%%%%%%%%%%%

\subsection{Ansatz and boundary conditions}
\label{PartAnsatzBCSingle}

According to our definition of single-field strings, we consider the case where
only the restricted representations of $\mathbf{\Sigma}$ and $\overline{ \mathbf{\Sigma}}$ are charged under $\text{U}(1)_{\text{str}}$. We normalize their charges to $q_\Sigma = 1$ and
$q_{\bar{\Sigma}}=-1$, the charges being defined by identifying $\tau_X=q_X
\text{Id}_X$ for an abelian generator in Eq.~(\ref{CovDev}).
To describe the different sub-representations singlet under the SM, we use the decomposition of Eq. (\ref{SubMultiplet}), which gives
\begin{equation}
\label{VEVSigma}
\langle \mathbf{\Sigma}(r,\theta) \rangle = \sigma(r,\theta) {\langle \mathbf{\Sigma} \rangle}_0,
\end{equation}
and
\begin{equation}
\label{VEVSigmab}
\langle \overline{ \mathbf{\Sigma}}(r,\theta) \rangle = \bar{\sigma} (r,\theta){\langle
 \overline{ \mathbf{\Sigma}} \rangle}_0,
\end{equation}
with ${\langle \mathbf{\Sigma} \rangle}_0^\dagger{\langle
 \mathbf{\Sigma} \rangle}_0=1$ and ${\langle \overline{ \mathbf{\Sigma}} \rangle}_0={\langle
 \mathbf{\Sigma} \rangle}_0^\dagger$.
 
The $D$-term condition associated with $\tau^{\text{str}}$, i.e. \cite{Martin:1997ns}
\begin{equation}
\label{D-term}
D^{\text{str}}=-g\sum_X (X^\dagger \tau^{\text{str}}_X X)=0,
\end{equation}
ensures that $\sigma$ and $\bar{\sigma}$ have the same norm. In addition, as the phase of the inflaton $S$ have been rephased in order to make $M$ real, it ensures that the global minimum of the potential is reached when $\boldsymbol\Sigma \overline{\boldsymbol\Sigma}=M^2 \in \mathbb{R}$. Both these results impose that $\overline{\boldsymbol\Sigma}=\boldsymbol\Sigma^\dagger$, which finally gives $\overline{\sigma}=\sigma^*$.

 The string itself is defined through
\cite{Kibble:1976sj,Hindmarsh:1994re}
\begin{equation}
\label{null_center}
\langle \mathbf{\Sigma} \rangle_ {(r=0)} = \langle \overline{ \mathbf{\Sigma}} \rangle_
 {(r=0)} = 0.
\end{equation}
Thus, we can introduce an ansatz similar to the Nielsen-Olesen solution, with integer winding number $n$ : \cite{Aryal:1987sn,Peter:1992dw,Ma:1992ky,Hindmarsh:1994re}
\begin{equation}
\label{Antsatz}
\begin{array}{l}
\sigma=f(r)\text{e}^{in\theta},\\
\phi_{x,\alpha}=\phi_{x,\alpha}(r),\\
S=S(r),\\
A_\mu=A_\theta^{\text{str}}(r) \tau^{\text{str}} \delta_\mu^\theta,
\end{array}
\end{equation}
where $f(r)$ and $A_\theta^{\text{str}}(r)$ are real, $\phi_{x,\alpha}(r)$ and $S(r)$ being complex. At infinity, we have 
\begin{equation}
\label{BCdebut}
\begin{array}{l}
\displaystyle{\lim_{r \to \infty} f(r)=M},\\
\displaystyle{\lim_{r \to \infty} A_\theta^{\text{str}}(r)=\frac{n}{g}},
\end{array}
\end{equation}
i.e. the value of $f$ ensures Eq.~(\ref{SigmaInf}) holds, while
the gauge field cancels $D_\mu \mathbf{\Sigma}$. 
After generalizing the notation of Eq. (\ref{VEV1}), the boundary conditions for the other fields satisfy
\begin{equation}
\begin{array}{l}
\displaystyle{\lim_{r \to \infty} \phi_{x,\alpha}(r)=\phi_{x,\alpha,{\scriptscriptstyle{(+)}}},}\\
\displaystyle{\lim_{r \to \infty} S(r)=S_{\scriptscriptstyle{(+)}}}.
\end{array}
\end{equation}
At the center of the string,
\begin{equation}
f(0)=0, ~~~~~~~ \text{and} ~~~~~~~ A_\theta^{\text{str}}(0)=0,
\end{equation}
while cylindrical symmetry imposes
\begin{equation}
\label{BCfin}
\displaystyle{\frac{\text{d}\phi_{x,\alpha}}{\text{d} r}(0)=0},
 ~~~~~~~ \text{and} ~~~~~~~
\displaystyle{\frac{\text{d} S}{\text{d} r}(0)=0}.
\end{equation}

%%%%%%%%%%%%%%%%%%%%%%%%%%%%%%%%%%%%%%%%%%%%%%%%%%%%%%%%%%%%%%%%%%

\subsection{Lagrangian and equations of motion}
\label{LagEOM}

With the ansatz of the previous section, we can simplify the model tremendously. The kinetic term yields (with implicit summations on the representations singlet of the SM)
\begin{multline}
\label{KSingleField}
K = - 2 \left|(\nabla_\mu-ig A_\mu^{\text{str}})\sigma\right|^{2}
-(\nabla_\mu \phi_{\bar{\imath},\bar{\alpha}})^*(\nabla^\mu \phi_{\bar{\imath},\bar{\alpha}})
-(\nabla_\mu \phi_{i,\alpha})^*(\nabla^\mu \phi_{i,\alpha})\\
 -(\nabla_\mu \phi_{I,\alpha})^*(\nabla^\mu \phi_{I,\alpha})
 -\left| (\nabla_\mu S)\right|^2 - \frac{1}{4}F_{\mu \nu}^{\text{str}} F^{\mu \nu \, \text{str} },
\end{multline}
where $F_{\mu \nu}^{\text{str} } = \nabla_\mu A_\nu^{\text{str}} - \nabla_\nu A_\mu^{\text{str}}$.
We label $(\bar{\imath},\bar{\alpha})$ the restricted representation
complex conjugate to $(i,\alpha)$.
No scalar products
between vectors in representation spaces are present due to the
normalization choice of Sec.~\ref{HI&SSB}. There is no cross-terms since singlet quadratic terms can only be built from products of two conjugate representations.

Writing down the potential is a bit trickier, as it contains high order terms
whose derivation w.r.t. the fields need to be done with care.
We add a subscript to the VEV indicating in which
representation is the product we consider,
${\langle X Y\rangle}_{Z}$ denoting the field in the
representation of $Z$ coming from the contraction between $X$ and $Y$. 

In the SO(10) case, 
the \textbf{126} and
$\mathbf{\overline{126}}$ representations are fifth-rank anti-symmetric respectively
self-dual\footnote{Self-duality being here defined by $\Sigma_{ijklm}=\frac{i}{5!}\epsilon_{ijklmabcde}\Sigma_{abcde}$.} and anti-self-dual
tensors $\Sigma_{ijklm}$ and $\bar{\Sigma}_{ijklm}$, while the
\textbf{210} is a fourth-rank anti-symmetric tensor
$\Phi_{ijkl}$. The singlet which can be formed with these fields is
$\bar{\Sigma}_{ijklm}\Sigma_{ijkno}\Phi_{lmno}$. Differentiating
with respect to $\bar{ \mathbf{\Sigma}}$, we obtain $\langle\mathbf{\Sigma}\mathbf{\Phi}\rangle_{\Sigma}$ in the same
representation as $ \mathbf{\Sigma}$, which is $\frac{1}{2}\left(\Phi_{[ij|\alpha\beta}\Sigma_{\alpha\beta |klm]}+\frac{i}{5!}\epsilon_{ijklmabcde}\Phi_{ab\alpha\beta}\Sigma_{\alpha\beta cde}\right)$
\cite{Slansky:1981yr,Allys:2015kge}, totally antisymmetric and self-dual as expected.

With these notations and with implicit summation on all the indices but e.g. $I$ when considering $V_{\Phi_I}$, the scalar potential becomes (noting that there is no $\beta$-coupling, as explained in Sec.~\ref{PartTwoClasses})
\begin{equation}
\label{V_S}
V_S=\kappa^2 (\sigma \sigma^*-M^2)^2,
\end{equation}
for the inflaton part,
\begin{multline}
V_\Sigma= m_\Sigma^{2} \sigma\sigma^* + \kappa^2 S S^* \sigma
\sigma^* +|\eta_x|^2|{\langle
 \overline{ \mathbf{\Sigma}}\mathbf{\Phi}_{x,\alpha}\rangle}_{\bar{\Sigma},0}|^2\sigma\sigma^*\phi_{x,\alpha}\phi_{x,\alpha}^* +m_\Sigma \kappa \sigma\sigma^* S^* +
\text{h.c.}\\ +
m_\Sigma \eta_x^*{\langle \overline{ \mathbf{\Sigma}}\rangle}_0{\langle
 \overline{ \mathbf{\Sigma}}\mathbf{\Phi}_{x,\alpha}\rangle}_{\bar{\Sigma},0}^\dagger
\sigma\sigma^*\phi_{x,\alpha} +\text{h.c.} + \eta_x \kappa {\langle
 \overline{ \mathbf{\Sigma}}\mathbf{\Phi}_{x,\alpha}\rangle}_{\bar{\Sigma},0}{\langle
 \overline{ \mathbf{\Sigma}}\rangle}_0^\dagger \sigma \sigma^*S^*\phi_{x,\alpha} +
\text{h.c.},
\end{multline}
and
\begin{equation}
V_{\bar{\Sigma}}=V_\Sigma ( \mathbf{\Sigma} \longleftrightarrow \overline{\mathbf{\Sigma}} ),
\end{equation}
for the string forming field part, and 
\begin{multline}
\label{V_Phi}
V_{\Phi_I}=
|m_{IJ}|^2\phi_{J,\alpha}\phi_{J,\alpha}^*+|\eta_I|^2|\langle \mathbf{\Sigma}
\overline{ \mathbf{\Sigma}}\rangle _{\Phi_I,
 0}|^{2}(\sigma\sigma^*)^2 +|\lambda_{Ixy}|^2|{\langle\mathbf{\Phi}_{x,\alpha}\mathbf{\Phi}_{y,\beta}\rangle}_{\Phi_I,0}|^2\phi_{x,\alpha}\phi_{x,\alpha}^*\phi_{y,\beta}\phi_{y,\beta}^*\\
 +m_{IJ}\eta_I^*{\langle\mathbf{\Phi}_{J,\alpha}\rangle}_0{\langle
 \mathbf{\Sigma} \overline{ \mathbf{\Sigma}}\rangle}_{\Phi_{J,\alpha},0}^\dagger \sigma
\sigma^* \phi_{J,\alpha}+\text{h.c.} +
m_{IJ}\lambda_{Ixy}^*{\langle\mathbf{\Phi}_{J,\gamma}\rangle}_0{\langle\mathbf{\Phi}_{x,\alpha}\mathbf{\Phi}_{y,\beta}\rangle}_{\Phi_{J,\gamma},0}^\dagger
\phi_{J,\gamma}\phi_{x,\alpha}^*\phi_{y,\beta}^* + \text{h.c.} \\ +
\eta_I \lambda_{Ixy}^*{\langle \mathbf{\Sigma}
 \overline{ \mathbf{\Sigma}}\rangle}_{\Phi_I,0}{\langle\mathbf{\Phi}_{x,\alpha}\mathbf{\Phi}_{y\beta}\rangle}_{\Phi_I,0}^
\dagger \sigma \sigma^* \phi_{x,\alpha}^*\phi_{y,\beta}^* +
\text{h.c.},
\end{multline}
\begin{equation}
V_{\Phi_i}=V_{\Phi_I} ( I \longleftrightarrow i, J \longleftrightarrow \bar{\jmath}),
\end{equation}
\begin{equation}
V_{\Phi_{\bar{\imath}}}=V_{\Phi_I} ( I \longleftrightarrow \bar{\imath}, J \longleftrightarrow j),
\end{equation}
for the other fields. An example of such potential with the conventions used in this paper can be found in Ref.~\cite{Allys:2015kge}.

The Lagrangian is
\begin{equation}
\label{Lagrangian}
\mathcal{L}= K - V_S - V_\Sigma - V_{\bar{\Sigma}} -
\sum_{x=I,i,\bar{\imath}}V_{\Phi_x},
\end{equation}
with the kinetic term given in Eq. (\ref{KSingleField}). Note that we recover the abelian Higgs model with no additional numerical factor in the limit described in Sec.~\ref{ToyModel}, due to the conventions used.

Finally, using the ansatz given in Eq. (\ref{Antsatz}), we obtain
the following equations of motion
\begin{equation}
\label{EOM}
\begin{array}{l}
\displaystyle{2\left(f''+\frac{f'}{r}\right)=\frac{fQ^2}{r^2}+\frac{1}{2}\frac{\partial
 V}{\partial f}},\\
  \displaystyle{\phi_{I,\alpha}''+\frac{\phi_{I,\alpha}'}{r}=\frac{\partial
 V}{\partial \phi_{I,\alpha}^*}},\\
  \displaystyle{\phi_{i,\alpha}''+\frac{\phi_{i,\alpha}'}{r}=\frac{\partial
 V}{\partial \phi_{i,\alpha}^*}},\\ 
 \displaystyle{\phi_{\bar{\imath},\bar{\alpha}}''+\frac{\phi_{\bar{\imath},\bar{\alpha}}'}{r}=\frac{\partial V}{\partial \phi_{\bar{\imath},\bar{\alpha}}^*}},\\
  \displaystyle{S''+\frac{S'}{r}=\frac{\partial V}{\partial S^*}},\\ 
  \displaystyle{Q''-\frac{Q'}{r}=2 g^2 f^2Q},
\end{array}
\end{equation}
where a prime means a derivative with respect to the radial
coordinate ($'\equiv \text{d}/\text{d}r$). We also introduced the field
\begin{equation}
\label{defQ}
Q(r)=n-gA_\theta^{\text{str}}(r),
\end{equation}
which is a real field function of $r$ only with for boundary
conditions
\begin{equation}
\label{CLQ}
Q(0)=n, ~~~~~~~ \text{and} ~~~~~~~ \displaystyle{\lim_{r \to \infty} Q(r)=0}.
\end{equation}

Eq. (\ref{EOM}) with associated boundary conditions can only be solved once the actual theory is implemented, thus giving the relevant and necessary coefficients in the potential.

%%%%%%%%%%%%%%%%%%%%%%%%%%%%%%%%%%%%%%%%%%%%%%%%%%%%%%%%%%%%%%%%%%
\subsection{Modification of the energy per unit length}
\label{OdGsingle}

We now evaluate the influence of the extra fields contribution to the energy per unit length, when a perturbative study of the condensation of these extra fields in the core of string is possible.

We take into account the possibility to work with high dimensional representations, of characteristic dimension $N$, as is often the case in GUTs. Let us remind that we chose the vectors which define the VEV directions to be normalized, i.e.
$\langle \mathbf{\Phi}_{x,\alpha}\rangle_0 \langle
\mathbf{\Phi}_{x,\alpha}\rangle_0^\dagger=1$ and $\langle \mathbf{\Sigma} \rangle_0
\langle \mathbf{\Sigma}\rangle_0^\dagger=1$. The cubic and
quartic contraction between these VEV directions must be smaller than
$1$ due to the Cauchy-Schwarz inequality. They can be approximately
estimated\footnote{For 
this purpose, we describe the VEV directions by vectors of $N$ 
components of value $1/\sqrt{N}$ (in order to be normalized), and the vectors 
formed from two different VEV directions by $N$ component of values $1/N$. 
A scalar product of two vectors giving $N$ times the product of their components, 
we found the results used.}
 to be of order $1/\sqrt{N}$ for the cubic terms, for example
${\langle\mathbf{\Phi}_{x,\alpha}\mathbf{\Phi}_{y,\beta}\rangle}_{\Phi_{z,\gamma},0}
\langle \mathbf{\Phi}_{z,\gamma}\rangle _0^\dagger$, and of order $1/N$ for the
quartic terms, as ${\langle \mathbf{\Sigma}
 \overline{ \mathbf{\Sigma}}\rangle}_{\Phi_I,0}{\langle\mathbf{\Phi}_{x,\alpha}\mathbf{\Phi}_{y\beta}\rangle}_{\Phi_I,0}^
\dagger $.

Important remarks must be done at this step. On the one hand, we consider here a dynamical case, and not only a static and uniform configuration, where it is sufficient to independently ensure that the different $F$-terms vanish (e.g. when studying the SSB scheme). Then, the different multiplicities of the $F$-term components must be taken into account. On the other hand, the choice of conventions in the definition of the superpotential and the kinetic part of the Lagrangian (where we included here no multiplicative factor) and of normalizations for the constant vectors in the representation space (see Sec.~\ref{RestrictedRep}) can considerably affect the formulation of the model. 

A complete model using the conventions of the present paper is described in Ref.~\cite{Allys:2015kge}. In this SO(10) case, taking the VEV directions for the singlets of the SM as given in
Ref. \cite{Bajc:2004xe,Aulakh:2003kg,Fukuyama:2004xs}, and after normalization, we find cubic
coefficients like $1/(10\sqrt{2})$ and $1/(6\sqrt{6})$, and quartic
coefficients like $1/54$ and $1/164$. This is roughly in agreement with
what was expected with a characteristic dimension of order $100$ for
the representation, and sufficient to do a first approximation. Note that Ref.~\cite{Allys:2015kge} also shows that the different formulations are indeed equivalent.

Let us evaluate a rough order of magnitude for the potential, and for that purpose consider a generic field $\phi$, without specifying
indices. From Eq. (\ref{V_Phi}), we have
\begin{equation}
V \simeq m^2\phi^2+ \frac{\eta^2 \sigma^4}{N}+ \frac{\lambda^2 \phi^4}{N} 
+\frac{m \eta \phi \sigma^2}{\sqrt{N}}+ \frac{m \lambda \phi^3}{\sqrt{N}}+\frac{\eta \lambda \sigma^2 \phi^2}{N}.
\end{equation}
To evaluate the contributions of $\phi$ to the energy, we consider its characteristic scale of variation due to the presence of the string. For this purpose, we estimate the value of $\phi$ at the center of the string and at infinity by taking the values which minimize the potential for $\sigma=0$ and $\sigma=M$, that we write respectively $\phi_0$ and $\phi_0 + \phi_1$. We obtain $\phi_0 \sim (\sqrt{N} m)/\lambda$, and $\phi_1 \sim(\eta M^2)/(\sqrt{N} m)$ for the perturbation parameter, leading to
\begin{equation}
\frac{\phi_1}{\phi_0}\sim \frac{\lambda \eta M^2}{N m^2}.
\end{equation}
Since $N\gg 1$, $M\leq m$, and one has $\lambda$ and $\eta \leq 1$ in most of the cases, a first approach by considering the modifications of $\phi$ as a perturbation is often relevant. 

When $\phi_1 \ll \phi_0$, we can evaluate the maximal modification of the energy per unit length due to the condensation of $\phi$ in the string. Close to the configuration with $\phi_0$, which is a minimum of the potential, the additional potential term is
\begin{equation}
\label{dU1}
\delta U \simeq \int r \text{d}r ~ m^2 \phi_1^2\sim \frac{\eta^2 M^2}{N \kappa^2 },
\end{equation}
which finally gives, since $U_0 \sim M^2$,
\begin{equation}
\label{dU1/U}
\frac{\delta U}{U_0} \sim \frac{\eta^2}{N \kappa^2}.
\end{equation}
This result gives a criterion to estimate if the toy model description of the cosmic strings is relevant from a macroscopic point of view. Note that when $\phi_1$ can not be treated as a perturbation or when $\delta U \ll U$ is not verified, the estimate (\ref{dU1}) of $\delta U$ is meaningless, and a complete calculation must be done.

Up to now, we did not consider the inflaton $S$. A careful examination however shows that we can estimate the modifications of $S$ due to presence of the string to be of order $\phi_1$, and that the contribution to the energy per unit length of this field is of the same order or lower than the contribution of $\phi_1$. It can be understood by the fact that $S$ has no characteristic scale, which is necessary for it in order to play the role of the inflaton.

%%%%%%%%%%%%%%%%%%%%%%%%%%%%%%%%%%%%%%%%%%%%%%%%%%%%%%%%%%%%%%%%%%

%%%%%%%%%%%%%%%%%%%%%%%%%%%%%%%%%%%%%%%%%%%%%%%%%%%%%%%%%%%%%%%%%%

\section{Many-field strings}
\label{ManyFieldsStrings}

\subsection{Ansatz and boundary conditions}
\label{IntroManyFields}

We now turn to the case of a many-field string assuming the minimal structure ansatz, i.e. with only the restricted representations singlet under the SM taking non vanishing values. We denote by $\tilde{\mathbf{\Phi}}_{x,\alpha}$ the sub-representations charged under U(1)$_{\text{str}}$, using as before the notations of Sec.~\ref{RestrictedRep} for the different restricted representations. In the case of a many-field string, such representations appear in the GUT field content, and the superpotential contains some $\beta$-couplings. As U(1)$_{\text{str}}$ is not broken before the last step of SSB, see Sec.~\ref{HI&SSB}, these particular fields must have vanishing VEVs before the end of inflation, yielding $\langle \tilde{\mathbf{\Phi}}_{x,\alpha}\rangle =0$.

An example of such a field can be the \textbf{16} representation of
SO(10), with a $\beta$ coupling through
\textbf{16}$\times$\textbf{16}$\times\mathbf{\overline{126}}$. Its
SM singlet is contained in its
($\mathbf{1},\mathbf{2},\mathbf{4}$) restricted
representation under the Pati-Salam group
[SU(2)$_L\times$SU(2)$_R\times$SU(4)]. Then, the $\beta$-coupling can be constructed from this previous sub-representation and the ($\mathbf{1},\mathbf{3},\overline{\mathbf{10}}$) representation under Pati-Salam contained in the $\overline{\mathbf{126}}$ representation of SO(10). Decomposing SU(4) to SU(3)$_C\times$U(1)$_{B-L}$, the restricted representation of this additional field is
charged under the U(1)$_{B-L}$ which can play the role of
U$(1)_\text{str}$ \cite{Slansky:1981yr,Jeannerot:2003qv}. A similar coupling can be formed with the $\overline{\mathbf{16}}$ and $\mathbf{126}$ representations.

More than one field and its conjugate are now charged
under U$(1)_\text{str}$, so we have to specify these charges. We
call them $q_{x,\alpha}$ and $q_\Sigma=-q_{\bar{\Sigma}}$. 
As we can multiply these charges (which can be fractional) by a common factor without loss of
generality, we choose the smallest set of charges where they all are
integers, i.e where a rotation of $2\pi$ in the U(1) group is the smallest one which reduces to identity.
%This definition ensures
%that a rotation through an angle $2 \pi$ in U$(1)_\text{str}$ (i.e. with a factor
%e$^{iq_X 2 \pi}$ for the field $X$) is the smallest one which leaves all
%the fields invariant. 
It will be convenient in what follows to define the winding number.

The $D$-term condition given in Eq. (\ref{D-term}) should now include all the fields that are charged under
U$(1)_\text{str}$. This condition is \emph{a priori} not sufficient to ensure that
$\sigma^{*}=\bar{\sigma}$ and similar relations for all the
fields charged under this abelian symmetry, i.e.
$\tilde{\phi}_{x,\alpha}^{*}=\tilde{\phi}_{\bar{x},\bar{\alpha}}$, with $\tilde{\mathbf{\Phi}}_{\bar{x},\bar{\alpha}}$ in the conjugate representation of $\tilde{\mathbf{\Phi}}_{x,\alpha}$. 
We will however assume it from now on, since this is a solution of $D_{\text{str}}=0$, and so a global minimum of the associated potential.

The same topological arguments than above imply that all the fields which
are charged under U$(1)_\text{str}$ vanish at the center of
the string \cite{Hindmarsh:1994re}:
\begin{equation}
\left\{
\begin{array}{l}
\langle \mathbf{\Sigma} \rangle_ {(r=0)} = \langle \overline{ \mathbf{\Sigma}} \rangle_
 {(r=0)} = 0,\\ \langle \tilde{\mathbf{\Phi}}_{x,\alpha}\rangle_
 {(r=0)}=0.
\end{array}
\right.
\end{equation}

We consider the following ansatz for an abelian cosmic string associated with the
generator $\tau^{\text{str}}$,
\begin{equation}
\label{Antsatz2}
\begin{array}{l}
\sigma=f_\sigma (r)\text{e}^{iq_\Sigma n\theta},\\
\tilde{\phi}_{x,\alpha}=f_{x,\alpha}(r)\text{e}^{iq_{x,\alpha} n(\theta-\theta_{x,\alpha})},\\
\phi_{x,\alpha}=\phi_{x,\alpha}(r),\\
S=S(r),\\
A_\mu=A_\theta^{\text{str}}(r) \tau^{\text{str}} \delta_\mu^\theta.
\end{array}
\end{equation}
In these equations, the $f_X$ and $A_\theta^{\text{str}}$ are real functions, and
$\theta_{x,\alpha}$ are reals constants. The possibility to freely
define an origin for the coordinate $\theta$ permits to put no initial
phase to the field $\sigma$. The integer $n$ is well identified with the
winding number, taking into account the choice of normalization we did
for the charges.
We also have the following boundary conditions, similarly to the single-field string case, 
\begin{equation}
\label{CLinf}
\begin{array}{l}
\displaystyle{\lim_{r \to \infty} f(r)=M,}\\
\displaystyle{\lim_{r \to \infty} f_{x,\alpha}(r)=|\tilde{\phi}_{x,\alpha,{\scriptscriptstyle{(+)}}}|,}\\
\displaystyle{\lim_{r \to \infty} \phi_{x,\alpha}(r)=\phi_{x,\alpha,{\scriptscriptstyle{(+)}}},}\\
\displaystyle{\lim_{r \to \infty} S(r)=S_{\scriptscriptstyle{(+)}},}\\
\displaystyle{\lim_{r \to \infty} A_\theta^{\text{str}}(r)=\frac{n}{g}},
\end{array}
\end{equation}
at infinity, and 
\begin{equation}
\begin{array}{l}
f(0)=0,~~~~~~
\tilde{\phi}_{x,\alpha}(0)=0, ~~~~~
 \displaystyle{A_\theta^{\text{str}}(0)=0},\\
\displaystyle{\frac{\text{d}\phi_{x,\alpha}}{\text{d} r}(0)=0},
 ~~~~~~~ \text{and} ~~~~~~~
\displaystyle{\frac{\text{d} S}{\text{d} r}(0)=0},
\end{array}
\end{equation}
at the center of the string.

At infinity, our ansatz simply gives an absolute minimum for the
potential on which we applied a local U$(1)_\text{str}$ transformation. It is the standard restated result for a cosmic strings,
but here with several fields charged under the string forming U(1). 
It shows that the angles $\theta_{x,\alpha}$ are not freely
chosen, but are those which give the absolute minimum of the potential and thus
define the SM symmetry at infinity, in addition to the limits
given in Eq. (\ref{CLinf}).

%%%%%%%%%%%%%%%%%%%%%%%%%%%%%%%%%%%%%%%%%%%%%%%%%%%%%%%%%%%%%%%%%%

\subsection{Modification of the energy per unit length}
\label{ManyFieldsModifEnergy}

We now turn to the modification of the energy per unit length from standard toy models due to the condensation of the additional fields in the core of the string, with a perturbative approach. As in the case of a single-field string, we work with a generic field $\phi$, without considering its indices anymore. 

The potential term gives
\begin{align}
V \simeq& ~ m^2\phi^2+ \frac{\eta^2 \sigma^4}{N} + \frac{\beta^2 \sigma^2 \phi^2}{N}+ \frac{\lambda^2 \phi^4}{N} 
+\frac{m \eta \phi \sigma^2}{\sqrt{N}} + \frac{m \beta \sigma \phi^2}{\sqrt{N}} \nonumber\\
& + \frac{m \lambda \phi^3}{\sqrt{N}}+\frac{\eta \beta \sigma^3 \phi}{N}+\frac{\eta \lambda \sigma^2 \phi^2}{N} + \frac{\beta \lambda \sigma \phi^3}{N}.
\end{align}
Introducing $\phi_0 \sim (\sqrt{N} m)/\lambda$, two perturbation scales appears, $\phi_1 \sim(\eta M^2)/(\sqrt{N} m)$ and $\phi_1^\prime \sim (\beta M)/\lambda$. So, in order to make a perturbative study of the string, the small parameters we have to consider are 
\begin{equation}
\frac{\phi_1}{\phi_0}\sim \frac{\lambda \eta M^2}{N m^2} ~~~~ \text{and} ~~~~\frac{\phi_1^\prime}{\phi_0}\sim \frac{\beta M}{\sqrt{N} m}.
\end{equation}
If a perturbative study is possible, we then estimate the maximal modification of $U$ due to the second term to be of order 
\begin{equation}
\delta U ^\prime \simeq \int r \text{d}r ~ m^2 \phi_1^2 \sim \frac{\beta^2 m^2}{\kappa^2 \lambda^2},
\end{equation}
which gives
\begin{equation}
\label{dU2/U}
\frac{\delta U^\prime}{U_0} \sim \frac{\beta ^2 m^2}{\kappa^2 \lambda^2 M^2}.
\end{equation}
We see that even in a model where the coupling constants are smaller than $1$, we cannot in general consider that the modification of the energy per unit length of the strings can be treated as a perturbation. It means that in the case of a many-field string, it is necessary to do a complete study of the microscopic structure of the string, and also that the toy model approximation is in general not valid.

Finally, we see that the simplification of Sec. \ref{Superpotential} consisting in
omitting terms like $\mathbf{\Sigma}\mathbf{\Sigma}\mathbf{\Phi}$ and
$\mathbf{\overline{\Sigma}}\mathbf{\overline{\Sigma}}\mathbf{\Phi}$ can be justified 
\emph{a posteriori}. Indeed, these terms must be studied in the case of 
a many-field string, as the field $\mathbf{\Phi}$ appearing in this cubic term 
has to be charged under U(1)$_{\text{str}}$ in order to provide a singlet term.
However, its contribution to the energy will be similar to a standard $\mathbf{\Sigma}
\mathbf{\overline{\Sigma}}\mathbf{\Phi}$ term, which is already taken into account here.

%%%%%%%%%%%%%%%%%%%%%%%%%%%%%%%%%%%%%%%%%%%%%%%%%%%%%%%%%%%%%%%%%%

\section{Conclusions and discussions}

Let us summarize by stating the different results obtained in this paper, focusing on the abelian strings forming at the last step of the SSB scheme, at the end of inflation. We worked in the framework of a SUSY GUT with hybrid inflation, but some results are more general, and we state for each point which hypotheses are actually necessary.
\begin{itemize}

\item[i)] The minimal structure model for realistic cosmic strings contains at least all the fields taking non vanishing VEVs at the end of the SSB down to the SM, which all condense in the core of the strings. These fields are singlets of the SM. This property neither depends on the GUT nor on the inflationary process.

\item[ii)] It is possible to distinguish two classes of strings, the single-field strings where only one field and its conjugate are charged under U(1)$_\text{str}$, and the many-field strings. The way the field content performs the SSB scheme is sufficient to determine which kind of strings form, independently of having a SUSY model or of the inflationary process. Different couplings appear in each model, which may give different phenomenologies for the strings.

\item[iii)] For each class of strings and with a SUSY GUT and hybrid inflation, we gave an ansatz and the associated boundary conditions to describe the minimal structure of the strings, completely defining it from a mathematical point of view. These strings are singlet of the SM. The fields take intermediate values in the core of the string between the minimum of the potential defining the SM at infinity and the configuration taken by the fields before the end of inflation. This ansatz can easily be generalized for other models.

\item[iv)] Going from the GUT description to the minimal structure ansatz, important numerical factors appear and must be taken into account. How these factors appear depends on the conventions and normalizations used. A special care has to be taken in SUSY models, where these factors are omitted when studying the static configurations which minimize the potential, i.e. with all $F$-terms independently vanishing. We emphasized how to normalize the GUT in order to recover the abelian Higgs model in the limit where the couplings between the string-forming Higgs and the other fields go to zero.

\item[v)] In the case of a SUSY GUT with hybrid inflation, we performed perturbative estimates of the modification of the energy per unit length with respect to standard toy models, which are given in Eqs.~(\ref{dU1/U}) and~(\ref{dU2/U}). These results are very different in the case of a single-field string, where the modifications are sizable in a high-coupling limit, and in the case of a many-field string, where the modifications are always important and require a complete computation of each model.

\end{itemize}

These results represent a first step of a more thorough investigation of cosmic strings taking into account their realistic structure. A complete study of the microscopic structure and the energy per unit length of such strings in a given SO(10) model has been performed in parallel, including numerical solutions, and can be found in Ref.~\cite{Allys:2015kge}. As discussed in Sec.~\ref{PartMinimalStructure}, several properties of the strings can also be modified in this framework and have major cosmological consequences. Indeed, the Higgs fields condensing in the core of the string could carry bosonic currents \cite{Witten:1984eb,Peter:1992dw,Peter:1992nz,Peter:1993tm,Morris:1995wd}. Moreover, their superpartner could carry fermionic currents through their zero modes \cite{Witten:1984eb,Jackiw:1981ee,Weinberg:1981eu,Davis:1995kk,Ringeval:2000kz,Peter:2000sw}. Furthermore, this complex microscopic structure could qualitatively modify the intercommutation process \cite{Shellard:1988ki,Laguna:1990it,Matzner:1988ky,Moriarty:1988qs,Moriarty:1988em,Shellard:1987bv} and thus the evolution of the cosmological string network, hence modifying its consequences on the CMB \cite{Kibble:1976sj,Kibble:1980mv,Copeland:1991kz,Austin:1993rg,Martins:1996jp,Martins:2000cs}. The cusps evaporation \cite{Srednicki:1986xg,Brandenberger:1986vj,Gill:1994ic,Olum:1998ag,Bhattacharjee:1989vu} also deserves more thoughts, since in our case, knowledge of the fields present in the string core implies knowledge of the relevant branching ratios into specific particles.

%ref courant fermioniques : \cite{Witten:1984eb,Jackiw:1981ee,Weinberg:1981eu,Davis:1995kk,Ringeval:2000kz,Peter:2000sw}
%We tried to keep the discussion as general as possible, while developing another parallel work devoted to the specific case of SO(10); this can be found in Ref.~\cite{Allys:2015kge}. Other aspects of the phenomenology of cosmic strings should also be studied in this realistic context, e.g. the formation of currents in the core of the string \cite{Witten:1984eb,Peter:1992dw,Peter:1992nz,Peter:1993tm,Morris:1995wd}. The consequences of this realistic structure regarding the processes of intercommutation \cite{Shellard:1988ki,Laguna:1990it,Matzner:1988ky,Moriarty:1988qs,Moriarty:1988em,Shellard:1987bv} should also be investigated, since it could modify the string network evolution and thus the observational consequences on the CMB \cite{Kibble:1976sj,Kibble:1980mv,Copeland:1991kz,Austin:1993rg,Martins:1996jp,Martins:2000cs}. In addition, the link done between the fields in the core of the string and the particle physics models could permit to investigate in more details the phenomenon of cusp evaporation \cite{Srednicki:1986xg,Brandenberger:1986vj,Gill:1994ic,Olum:1998ag,Bhattacharjee:1989vu}. Finally, the formation of non abelian vortices \cite{Aryal:1987sn,Ma:1992ky,Davis:1996sp} would need investigation to complete this general study.

%%%%%%%%%%%%%%%%%%%%%%%%%%%%%%%%%%%%%%%%%%%%%%%%%%%%%%%%%%%%%%%%%%

\subsection*{Acknowledgment}

I wish to thank Patrick Peter for many valuable discussions and suggestions, and also for a critical reading of the manuscript. I also thank R. Brandenberger and M. Sakellariadou for enlightening discussions and remarks.

\bibliographystyle{unsrt}

\end{document}